\documentclass[a4paper,11pt]{article}
\pdfoutput=1
\usepackage{jstyle}
\usepackage{color}
\usepackage{latexsym}
\usepackage{amssymb,comment}
\usepackage{shadow}

\def\sideremark#1{\ifvmode\leavevmode\fi\vadjust{\vbox to0pt{\vss
 \hbox to 0pt{\hskip\hsize\hskip1em
 \vbox{\hsize3cm\tiny\raggedright\pretolerance10000
 \noindent #1\hfill}\hss}\vbox to8pt{\vfil}\vss}}}%

\def\a{\alpha}
\def\b{\beta}
\def\g{\gamma}

\def\d{\delta}

\def\m{\mu}

\def\X{\Xi}

\def\s{\sigma}

\def\t{\tau}

\def\cC{{\cal C}}
\def\cD{{\cal D}}

\def\cG{{\cal G}}

\def\cL{{\cal L}}
\def\cM{{\cal M}}

\renewcommand{\X}{{\mathcal X}}
\newcommand{\Y}{{\mathcal Y}}
\newcommand{\Z}{{\mathcal Z}}
\renewcommand{\colon}{\scalebox{1.2}{:}}
\newcommand{\wD}{\widehat D}

\author[a]{Euihun Joung,}
\affiliation[a]{Scuola Normale Superiore and INFN,
Piazza dei Cavalieri 7, 56126 Pisa, Italy}
\emailAdd{euihun.joung@sns.it}

\author[b]{Massimo Taronna}
\affiliation[b]{Max Planck Institute for Gravitational Physics (Albert Einstein Institute)
Am M\"uhlenberg 1, D-14476 Golm, Germany}
\emailAdd{massimo.taronna@aei.mpg.de}

\author[c]{\& Andrew Waldron}
\affiliation[c]{Department of Mathematics, University of California Davis,
CA95616, USA}
\emailAdd{wally@math.ucdavis.edu}

\title{\centering A Calculus for Higher Spin Interactions}

\abstract{Higher spin theories can be  efficiently described in terms
of auxiliary St\"uckel\-berg or projective space field multiplets.
By considering how higher spin models couple to scale, these approaches
can be unified in a conformal geometry/tractor calculus  framework. We review these methods and apply  them
 to higher spin vertices to obtain a generating function for massless, massive and partially 
massless three-point interactions.}

\begin{document}

\maketitle

\section{Introduction}\label{sec: intro}

Massless and massive higher-spin interactions are believed to be governed by Vasiliev's
equations~\cite{Vasiliev:1988sa} and by String Theory, respectively; relating these theories is a pressing problem for modern theoretical physics. For the former, scattering amplitudes are formulated as ($d=2$) CFT vertex operator correlators while the Vasiliev system
relies on an unfolded frame-like approach. However, in the end, 
one is often interested in either an~$S$-matrix or Witten-type
diagrams, 
whose  features can often be determined by gauge invariance alone.
In a light-cone framework and flat backgrounds, detailed results for massless~\cite{Bengtsson:1983pd,Metsaev:1993ap,Fradkin:1995xy} and massive~\cite{Metsaev:2005ar,Metsaev:2007rn} cubic higher spin interactions were obtained by following exactly this philosophy. More recently, a covariant version of this program was carried out for higher spin cubic vertices, both for simple cases (see, for example~\cite{Zinoviev:2008ck,Boulanger:2008tg,Bekaert:2010hp,Manvelyan:2010wp,Zinoviev:2010cr,Henneaux:2012wg} and the review~\cite{Bekaert:2010hw}) and rather generally~\cite{Manvelyan:2010jr,Taronna:2010qq,Sagnotti:2010at,Fotopoulos:2010ay}. The (anti) de Sitter [(A)dS] and general mass (including partially massless [PM]) cases were then
given in~\cite{Joung:2011ww,Joung:2012rv,Joung:2012fv,Taronna:2012gb,Manvelyan:2012ww,Joung:2012hz}, while frame-like and mixed symmetry analyses were performed in~\cite{Vasilev:2011xf,Boulanger:2012dx} and~\cite{Alkalaev:2010af,Boulanger:2011qt,Boulanger:2011se,Lopez:2012pr}, respectively.  Early results beyond cubic order are available in both light-cone formalism~\cite{Metsaev:1991mt,Metsaev:1991nb} and covariant settings~\cite{Taronna:2011kt,Dempster:2012vw,Boulanger:2013zza}.

A central difficulty faced by higher spin theories is maintaining correct degrees of freedom (DoF) counts in the presence of interactions which generically destroy the gauge invariances or constraints controlling the DoF of free higher spin wave equations. For non-interacting theories, by including St\"uckelberg auxiliary fields, gauge invariance can be used as the central principle underlying the propagating higher spin DoF for all mass types: There are various ways to understand the auxiliary field content required for massive higher spin fields, crucial among them being their origin as Scherk--Schwarz reductions~\cite{Scherk:1979zr} of massless higher spins in one higher dimension~\cite{Rindani:1985pi,Aragone:1988yx}. Indeed, by a radial reduction corresponding to a conformal isometry of a flat embedding space~\cite{Biswas:2002nk}, the same mechanism generates the St\"uckelberg couplings for higher spins in constant curvature backgrounds~\cite{Hallowell:2005np}. This is the first hint that conformal geometry might play a {\it r\^ole} in these constructions. It also suggests an underlying Dirac space construction, where conformally
flat spaces are realized as sections of a cone in two higher dimensions. What is surprising is that such methods, which have long been known to be applicable to models with conformal symmetries~\cite{Dirac:1936fq}, can actually be used to great advantage for massive---non-conformal---models~\cite{Gover:2008pt,Gover:2008sw,Shaukat:2009hp}.

\begin{figure}
\begin{center}
 \includegraphics[scale=.20]{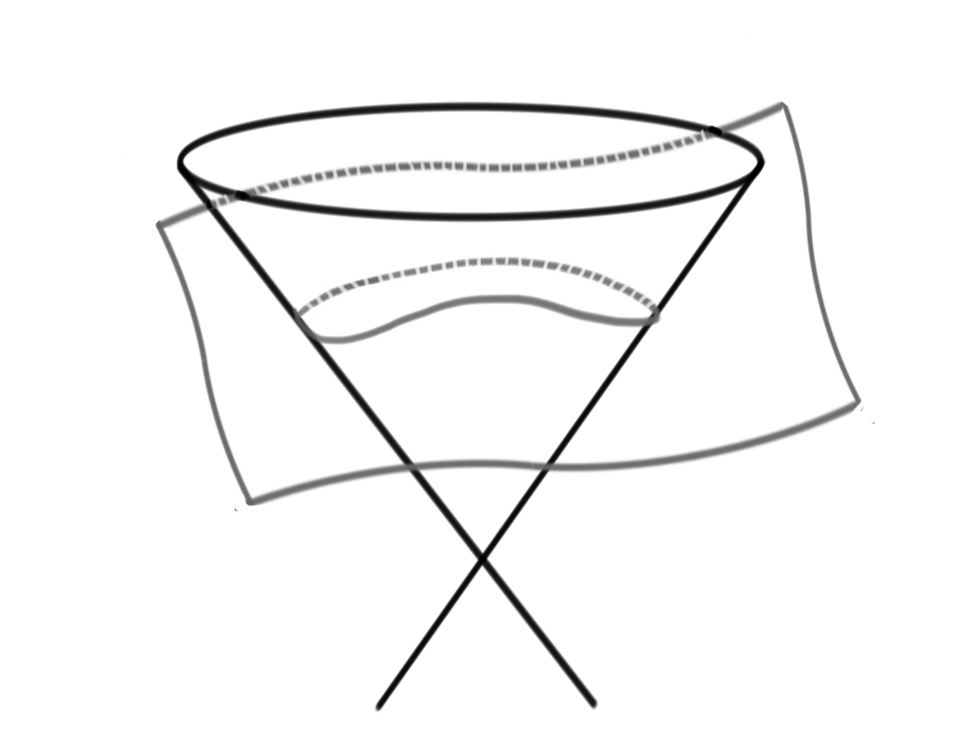}
\includegraphics[scale=.20]{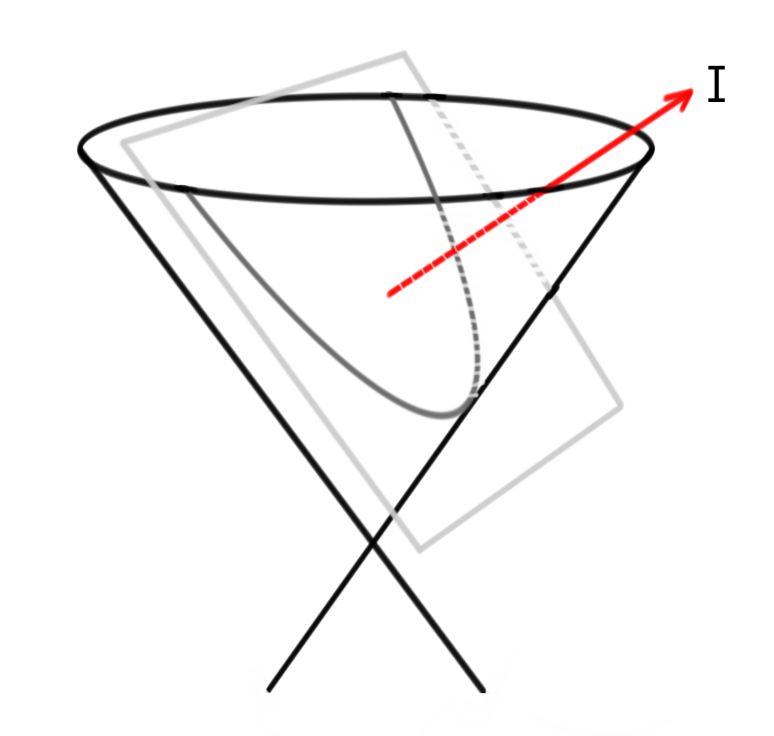} \hspace{5mm}
\end{center}
\caption{Conformally related metrics are obtained by slicing the conformal cone. These are conformally Einstein when there exist  slices admitting a parallel scale tractor~$I$. The second picture depicts the slice inducing an Einstein metric.\label{conepik}}
\end{figure}

The flat model for a~$d$-dimensional conformal geometry is obtained by sections of an ambient light-cone in~$(d+2)$-dimensions. Metrics induced on~$d$-dimensional slices by the~$(d+2)$-dimensional ambient metric are conformally related. Metrics induced by flat slices (the classical conic sections) give constant curvature spaces, as depicted in Figure~\ref{conepik}.
These are characterized by the normal vector~$I$---which we will later elevate to a parallel ambient vector field termed the scale tractor---to the flat slicing hypersurface.
Moreover this flat model can be generalized to the curved setting where the space of slicings yields general conformal classes of metrics,
while parallel scale tractors correspond to Einstein metrics. So in this picture, solving Einstein's equations amounts to finding parallel scale tractors~\cite{MR1322223}.

The relevance of a six dimensional cone to four dimensional conformal wave equations was first observed by Dirac~\cite{Dirac:1936fq} while its~$(d+2)$-dimensional curved generalization and application to~$d$-dimensional conformal geometry was initiated by Fefferman and Graham~\cite{MR837196}. The parallel scale tractor description of conformally Einstein metrics was discovered by Bailey, Eastwood and Gover in a paper which also developed the so-called ($d$-dimensional) ``tractor calculus'' for conformal invariants~\cite{MR1322223}. Later it was realized that tractors could also be profitably  described using ambient~$(d+2)$-dimensional tensors \cite{Cap:2002aj,Gover:2004as}. Moreover, it was shown that tractors could be used to express the fundamental wave equations of physics~\cite{Gover:2008pt,Gover:2008sw,Shaukat:2009hp}. The main idea was very simple: while parallel scale tractors~$I$ describe the background Einstein geometry, evolving boundary data along~$I$
corresponds to wave equations. This development allowed both massless and massive wave equations to be described by conformal geometry, rather than Riemannian geometry methods. Mass then amounts to how physical fields respond to changes of scale ({{\it i.e.}}, their tractorial weights).

\begin{figure}
\begin{center}
\includegraphics[scale=.24]{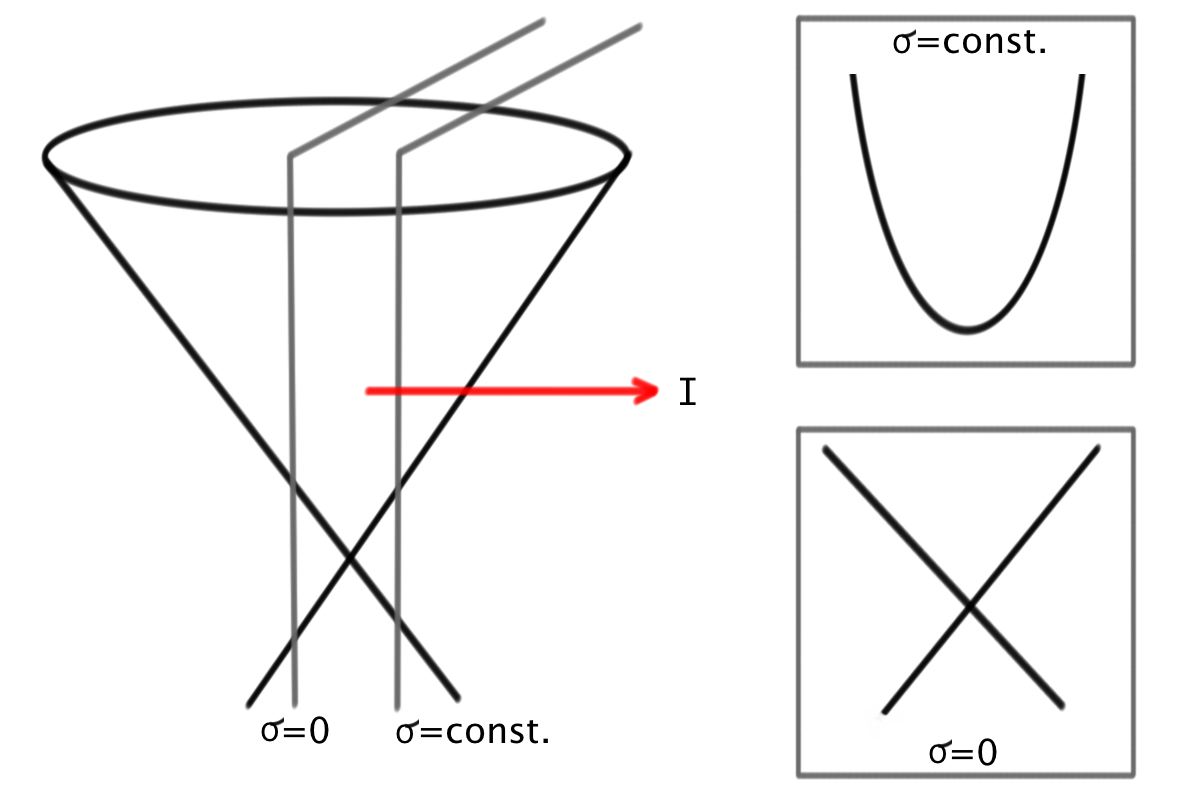}
\end{center}
\caption{\label{Evolution} Coupling to scale through the scale tractor~$I$ determines the evolution of physical fields
with masses labeled by conformal weights. The scale tractor also determines how data is moved from the boundary
(the zero scale slice) to the bulk (standardly described by a constant scale slice).}
\end{figure}

The above picture  becomes much richer when one considers also boundary problems, in particular those with data at conformal infinities. In fact, this is
precisely the setting of the AdS/CFT correspondence~\cite{Gover:2011rz,RodGover:2012ib}. Firstly, the slicing hypersurface is described by  the constant locus of a unit homogeneity scalar called the scale~$\sigma$, that plays the {\it r\^ole} of a dilaton field, or in other words a~$d$-dimensional scalar, conformal density. A key insight of Gover~\cite{MR2587388}, was that although a nowhere-vanishing scale and a conformal class of metrics is equivalent to a Riemannian geometry, this is not the case when~$\sigma$ has a non-trivial zero-locus. This  led to a generalization called ``almost Riemannian geometry''. In hyperbolic settings the zero locus of the scale~$\sigma$ amounts to a conformal infinity. This is depicted in the conformally flat---conic sections---setting in Figure~\ref{Evolution}.
Observe that constant loci of~$\sigma$ intersect the cone along hyperboloids (positively curved constant curvature spaces) while the zero locus yields a cone, and in turn conformal structure, in one dimension less. The former intersection corresponds to the bulk manifold in an AdS/CFT correspondence while the latter yields the boundary conformal geometry (and in turn CFT).

The power of this approach is that the bulk conformal structure can be utilized to realize spectrum or solution generating symmetries~\cite{Gover:2011rz,RodGover:2012ib}: The contraction of the scale tractor~$I$ with a tractor analog of the gradient and Laplace operators (known as the Thomas-$D$ operator~\cite{MR1322223}) yields the so-called Laplace-Robin operator. This is a conformal version of the bulk Laplacian which continues smoothly to the boundary (even though it is at conformal infinity). Remarkably, this operator is a generator of an~${\frak sl}(2)$ solution-generating algebra valid on any curved manifold~\cite{Gover:2011rz}. This facilitates solutions to conformal infinity boundary problems. These results have a wide applicability, both to higher spin, bose, fermi, massless, massive and PM systems.
Hence, the main building blocks for a calculus for scattering problems taking full advantage of the bulk conformal structure are now available.
The next (and crucial) step is to describe higher spin vertices in this approach. In this article we show how this can be done for  totally symmetric higher spin fields. This requires a melding of known results for these vertices with tractor approaches to higher spin fields.

Before summarizing our results, we provide a brief guide to the Article. In Section~\ref{Tractors} we review the tractor calculus description of conformal geometry and of physical systems in terms of conformally invariant tractors coupled to scale. 
In Section~\ref{ths} we specialize these methods to higher spins, focusing on their on-shell description. The results in Section~\ref{abcd}
focus on how to write point-split on-shell amplitudes ({\it \`a la}~\cite{Taronna:2010qq,Sagnotti:2010at,Joung:2011ww,Joung:2012rv,Joung:2012fv,Taronna:2012gb,Joung:2012hz}) in terms of tractor multiplets  and are new.
In Section~\ref{Cubic} we apply our ``tractor higher spin Noether method'' to compute the three point vertex generating functions.
In the Appendices, we derive various key identities and connect our results with previous ones based on a~$(d+1)$-dimensional projective space approach~\cite{Joung:2011ww,Joung:2012rv,Joung:2012fv,Taronna:2012gb,Joung:2012hz}.

\subsection*{Summary of results}

Our  results for totally symmetric higher spins of arbitrary rank can be compactly expressed in terms of tractor generating functions
$
\Phi(x,U)
$
(where a~$(d+2)$-dimensional auxiliary vector~$U^M$ is used to keep track of tractor bundle valued indices--see Section~\ref{oshst}).
Vertex generating functions can be expressed in terms of the irreducible  set of operators
\be
\Y_{i}=\partial_{U_i}\cdot\wD_{i+1}\,,\qquad
\Z_{i}=\s^{-2}\,\partial_{U_{i-1}}\!\!\cdot\partial_{U_{i+1}}
\qquad [i\simeq i+3]\,,
\nonumber
\ee
built from the  Thomas-$D$ operator (see Section~\ref{abcd}):
\begin{equation*}\label{gen cub}
	S^{\sst (3)}\sim
	\int_{\sigma}\,\colon\,
	\s^{\sum_i\tau_i}\,C(\Y_i\,,\Z_i)\,\colon\,
	\Phi(X_{1},U_{1})\,\Phi(X_{2},U_{2})\,\Phi(X_{3},U_{3})\,
	\Big|_{\overset{x_i=x}{\sst U_{i}=0}}\, .
\end{equation*}
Here, the integration measure~$\int_\sigma$ is defined in~\eqref{measure}, the normal ordering is~$\sigma>\Y>\Z$ and the parameters~$\t_i$ are the twists of respective fields. Our punchline is a proof that the tractor gauge consistency condition--which amounts to (strictly) massless ($\t_{1}=2$) gauge transformations in a dual $(d+2)$-dimensional theory~\cite{Gover:2008pt,Gover:2008sw,Gover:2009vc}, 
gives a differential equation determining the function~$C$\,:
\be
\Big[\Y_3\partial_{\Z_2}-\Y_2\partial_{\Z_3}\,-\hat\g\left(\Y_3\partial_{\Y_3}-\Y_2\partial_{\Y_2}+\tfrac{\tau_2-\tau_3}2\right)\partial_{\Y_1}\Big]\,C(\,\Y_i\,,\,\Z_i\,)\,=\,0\,,
\label{eq}
\ee
where
\[
\hat \g=\tau_{2}+\tau_{3}-d+1-2\,\sum_i \Z_i\partial_{\Z_i}\,.
\]
This equation has already been solved in~\cite{Joung:2012hz}: For that one absorbs the factor~$-\,\hat\g$ into a  differential operator~$\hat \delta$ \begin{equation*}
\hat\delta\,=\,-\s^{d}\,\frac{d}{d\s}\,\s^{-d+1}\, .
\end{equation*}
Exactly the same operator arose in~\cite{Joung:2012fv} from a careful handling of a projective space delta function measure.
In these terms, the cubic coupling for three massless fields can be written~as
\be\label{BIGFORMAGGIO}
C(\,\Y_{i}\,,\,\Z_{i}\,)\,=\,e^{-\,\hat\d\,\cD}\,K(\,\Y_{i}\,,\,\cG\,)\,\big|_{\cG=\sum_i\Y_i\,\Z_i}\,,\nonumber
\ee
where~$\cD:=\big[\Z_1\partial_{\Y_2}\partial_{\Y_3}+\Z_1\Z_2\partial_{\Y_3}\partial_{\cG}+\text{cyclic}\big]+\Z_1\Z_2\Z_3\,\partial_{\cG}^2$ (see~\cite{Joung:2012fv,Joung:2012hz}) and~$K$ is an arbitrary polynomial function of four variables.
 
The same pattern arises also for generic massive and (partially-)massless couplings.  These correspond to various intersections of kernels of the differential operator appearing in Eq.~\eqref{eq}, and cyclic permutations thereof, as discussed in~\cite{Joung:2012hz}. In summary, we find that  the solutions in the  projective formalism of~\cite{Joung:2012hz} and the corresponding  tractor ones are related  simply by replacing the~$(d+1)$-dimensional integration with the standard (conformally invariant)~$d$-dimensional measure along with substitutions~$Y\rightarrow\Y$ and~$Z\rightarrow \Z$. In particular, the tractor approach gives an 
alternative proof of the~$\d$-function methods used in~\cite{Joung:2011ww,Joung:2012rv,Joung:2012fv,Taronna:2012gb,Joung:2012hz}.

\section{Tractors}\label{Tractors}

A conformal~$d$-manifold~$\cM$ is a manifold equipped with a conformal class of metrics~$[g_{\mu\nu}]=[\Omega^2\,g_{\mu\nu}]$.
The data~$(\cM,[g])$ determines the standard tractor bundle~${\cal T}\cM$ over~$\cM$, which can be viewed as a conformally invariant extension
of the tangent bundle~$T\cM$. This comes equipped with a canonical {\it tractor connection}~$\nabla^{\cal T}$. In simple (four dimensional-)terms,
tractors replace four-vectors (sections of~$T\cM$) by six-vector sections of~${\cal T}\cM$ in order to make Weyl invariance manifest.
Under changes of Weyl frame~$g_{\mu\nu}(x)\mapsto\Omega^2(x)\,g_{\mu\nu}(x)$, a \emph{standard tractor}~$V^M\in \Gamma{\cal T}\cM$ ($M=0,\ldots,d+1$) transforms as
$$
V^M:=\begin{pmatrix}V^+\\ V^m \\ V^-
\end{pmatrix} \longmapsto \begin{pmatrix}\Omega V^+\\ V^m + \Upsilon^m V^+\\ \Omega^{-1} \big( V^- - \Upsilon_\mu V^\mu +\frac12 \Upsilon^2 V^+\big)
\end{pmatrix}=:U^M{}_N V^N\, .
$$
Here~$\Upsilon_\mu:=\Omega^{-1}\partial_\mu \Omega$ and we have used the vielbein in the middle slot to flatten indices. The matrix~$U^M{}_N$ is~$SO(d+1,1)$-valued\footnote{All formul\ae \ presented here continue to any metric signature by letting~$d\to (q,d-q)$ and thus~$(d+1,1)\to (q+1,d-q+1)$.}.
The tractor connection acts on~$V^M$ as
\newcommand{\Rho}{{\mbox{\sf P}}}
\newcommand{\J}{{\mbox{\sf J}}}
$$
\nabla_\mu^{\cal T} V^M = \begin{pmatrix}
\partial_\mu V^+- V_\mu\\ \nabla_\mu V^m + e_\mu{}^m V^+ + \Rho_\mu{}^m V^- \\
\partial_\mu V_- - \Rho_\mu{}^m V_m
\end{pmatrix}\, ,
$$
and is the covariant derivative with respect to the change of Weyl frame given above.
On the right hand side of this formula,~$\nabla$ denotes the Levi-Civita connection and the Schouten tensor~$\Rho_{\mu\nu}$ is defined by the decomposition of the Riemann tensor
into its trace-free Weyl  plus trace pieces:
$$
R_{\mu\nu\rho\sigma}=W_{\mu\nu\rho\sigma}+g_{\mu\rho}\Rho_{\nu\sigma} -g_{\nu\rho}\Rho_{\mu\sigma} +g_{\nu\sigma}\Rho_{\mu\rho} -g_{\mu\sigma}\Rho_{\nu\rho} \, .
$$
To complete the tractor calculus we introduce \emph{weighted tractors}~$V^M\in \Gamma{\cal T}\cM[w]$ transforming as
$$
V^M \mapsto \Omega^w\,U^M{}_N V^N\, ,
$$
as well as a pair of tractor operators, of weights~$-1$ and~$+1$, respectively known as the \emph{Thomas-$D$ operator} and \emph{canonical tractor}:
$$
D^M:=\begin{pmatrix}w(d+2w-2) \\ (d+2w-2) \nabla^m\\ -\Delta -w \J \end{pmatrix}\quad \mbox{and}\qquad X^M=\begin{pmatrix}0\\0\\1\end{pmatrix} \, .
$$
These both act on (weighted) tractor(-tensor)s yielding tractor(-tensor)s, for this reason we  have dropped the (implicit) label~${\cal T}$ on the tractor connection, also~$\Delta:=g^{\mu\nu} \nabla_\mu\nabla_\nu$ and~$\J:=\Rho^\mu_\mu$. Importantly, for any conformal structure~$(\cM,[g_{\mu\nu}])$, these operators obey a null condition
$D_M D^M = 0 = X^M X_M\,$,
where indices are raised and lowered with the~$SO(d+1,1)$-invariant tractor metric~$V\cdot V':=V^+ V'{}^-+V^- V'{}^+ + V^m V'_m$\,.

Since the Thomas-$D$ operator unifies the Laplacian and gradient operators in a single tractor multiplet of operators, it
will play a crucial {\it r\^ole} in many computations. Let us gather together some of its key properties:
Firstly, it is null, in the sense:
$$
D_M D^M=0\, .
$$
However, since it is second order in derivatives, it does not obey a Leibniz rule. Nonetheless, an integration by parts formula does hold (with an unusual sign)
\begin{equation}\label{parts}
\int_{\cM}\! d^{d}x\,\sqrt{-g}\ V^M\,D_M U = \int_{\cM}\! d^{d}x\, \sqrt{-g}\   (D_M V^M)\,U\, ,
\end{equation}
for any tractors~$V^M$ and~$U$ (suppressing further indices such that the overall integrand is a scalar) of weights~$w_V$ and~$w_U$ subject to~$d+w_V+w_U-1=0$ (which ensures that the integrand is of zero weight). Moreover, the failure of the Leibniz property can be characterized as follows: Acting on any tractor with weight~$w\neq 1-\frac d2$ we first define
$$
\wD^M:=\frac1{d+2w-2}\, D^M\, .
$$
Then if ~$A$ and~$B$ are tractors of weight~$w_A$ and~$w_B$, respectively, the failure of the Leibniz rule is measured by the following identity 
\begin{equation}
\label{Leibniz}
\wD^M(AB) -(\wD^M A)B-A(\wD^M B) = -\frac{2}{d+2w_A+2w_B-2} \, X^M (\wD^N A) (\wD_N B)\, .
\end{equation}
This is easily verified by using the ambient formula for the Thomas-$D$ operator given in~\eqref{tom D} below, and is valid away from obvious poles at distinguished values of~$w_A,w_B$. 
This formula can be further simplified to an operator statement by introducing the weight operator~$h$ whose eigenvalue is~$d+2w$ acting on weight~$w$
tractors:
$$
\wD^M A - (\wD^{M}A)-A\,\wD^{M}=-\frac2h\, X^M (\wD^{N} A)\,\wD_{N}\, .
$$
From time to time, we will need the commutator between the Thomas-$D$ and canonical tractor operators:
\be\label{XDcomm}
\left[X^M\,,\,\wD^N\right]\,=\,\frac{2}{h}\,X^N\,\wD^M\,-\,\eta^{MN}\,,
\qquad
\eta_{MN}=\begin{pmatrix}0&0&1\\0&\eta_{mn}&0\\1&0&0\end{pmatrix}\,,
\ee
where $\eta_{MN}$ is the $SO(d,2)$-invariant metric.

Finally, on conformally Einstein manifolds, the Thomas-$D$ operator commutes with the scale tractor~$I_M=\wD_M\,\sigma$ (see Section \ref{sec:we}):
$$
\left[ D^M\,,\, I^N\right]\,=\,0\,,\qquad [ \wD^M\,,\, I^N ]\,=\,0\,,
$$
while it commutes with itself on flat conformal structures:
$$
[\,D^{ M}\,,\,D^{N}\,]=0\,,\qquad [\,\wD^{M}\,,\,\wD^{ N}\,]=0\,.
$$

\subsection{Ambient tractors}\label{ambient}

The bundle-theoretic description of tractors and their calculus is extremely useful for computations whose output is required in standard Riemannian geometry terms.
However, for many computations, an ambient description of tractors is very powerful. For that we first introduce a Fefferman--Graham ambient space\footnote{Originally Fefferman and Graham studied~$(d+2)$-dimensional Ricci-flat ambient spaces~\cite{MR837196}. Ricci flatness is not required here (it can be viewed as a choice of gauge for the geometry extending away from the Dirac cone), nonetheless, we still employ the name Fefferman--Graham ambient space even in its absence. Flat Fefferman--Graham spaces reproduce the Dirac cone construction.}. This is a~$(d+2)$-dimensional
space~$\widetilde \cM$ endowed with a metric obeying
$$
G_{MN}=\nabla_M X_N
$$
for some vector field~$X_N$. This condition  immediately implies~$G_{MN}=\frac12\nabla_M \partial_N X^2$. The function~$X^2$ is known as a homothetic potential or a defining function; its zero locus defines a curved version of the Dirac cone described and depicted in the Introduction. The transition to the underlying~$d$-dimensional conformal geometry is achieved via
reducing to the cone and then demanding a homogeneity condition with respect to the homothety~$X^M$. More precisely, tractor(-tensor)s are equivalence classes of ambient tensors on~$(\widetilde \cM,G_{MN})$
\be\label{cone}
T^{M_1\cdots M_s} \sim T^{M_1\cdots M_s} + X^2\, S^{M_1\cdots M_s}\, ,
\ee
(where the tensor~$S$ extends smoothly to the cone~$\{X^2=0\}$; spinor-tractors can be defined analogously~\cite{Shaukat:2009hp}) 
classified by weights~$w$
$$
X\cdot \nabla \, T^{M_1\cdots M_s}=w\, T^{M_1\cdots M_s}\, .
$$
The equivalence relation~\eqref{cone} is precisely that enjoyed by the lightlike physical excitations of a massless scalar field in a momentum basis. Therefore,
tractor operators can be derived by considering the momentum representation of the~$\mathfrak{so}(d+2,2)$ generators of the conformal group acting on a flat ambient space~\cite{Gover:2009vc}. Their generally curved counterparts follow by replacement of partial derivatives by covariant ones. Thus, acting on ambiently represented tractors, the Thomas-$D$ operator is given by the analog of a momentum space conformal boost
\begin{equation}\label{tom D}
D_M=(d+2\,X\cdot\nabla)\,\nabla_M-X_M\,\nabla_{}^{2}\,.
\end{equation}
This construction ensures that~$D_M$ respects the equivalence relation~\eqref{cone}.
\subsection{Wave equations}
\label{sec:we}

To describe the evolution of physical fields we must consider how they couple to scale. This problem is solved by first
considering gravity. To begin with, suppose we are given a double conformal class of a metric and scale~$[g_{\mu\nu},\sigma]=[\Omega^2\, g_{\mu\nu},\Omega\, \sigma]$.  From this we can
construct the \emph{scale tractor}
$
I_M=\frac1d\,  D_M \sigma=\begin{pmatrix}-\frac12(\Delta \sigma + \J\sigma),&n_m,&\sigma\end{pmatrix}
$,
where~$n_\mu:=\partial_\mu \sigma$. Requiring that~$I^M$ is tractor parallel
$$
\nabla^{\cal T}_\mu I^M=0\, , 
$$
ensures that~$g_{\mu\nu}$ is conformal to an Einstein metric, with the Einstein metric being achieved precisely in the choice of Weyl frame~$\sigma={\rm constant}$.
This is the mathematics behind the conic sections picture of Einstein geometries sketched in the Introduction. Moreover, since~$I^M$ is parallel for conformally Einstein metrics,
its square~$I^{2}=I^M I_M$ is constant; physically this is the cosmological constant. Note that the Einstein--Hilbert action in these terms is simply the conformally invariant expression
$
S[g,\sigma]=\int d^{d}x \sqrt{-g}\,  \sigma^{-d} I^2
$, so that  Einstein's equations amount to extremizing the magnitude of the scale tractor (a cosmological term is just the integral of the conformally invariant measure:~$\int d^{d}x\sqrt{-g}\,  \sigma^{-d}$.) In fact, taking the normal vector~$n_\mu=\partial_\mu\sigma$ to loci of constant~$\sigma$ as an independent field, then the pair~$(\sigma,n_\mu)$ can be viewed as a generalized lapse and shift and thus the parallel scale tractor equation yields a covariant extension of the ADM formalism.

Not only does the scale tractor control the geometry, it determines the evolution of physical fields. If~$\Phi^\bullet$ is any tractor tensor, the quantity~$D^M \Phi^\bullet$ is covariant under
Weyl transformations. Generally, wave equations are not conformally invariant, so they must somehow be coupled to scale. There is a simple universal prescription for this, namely
the contraction with the scale tractor
$$
I_MD^M \Phi^\bullet=0\, .
$$
The operator~$I\cdot D:=I_M D^M$ is called the Laplace-Robin operator because in the bulk it is a conformally invariant version of the Laplacian while along the boundary
it gives the Robin operator, which is a conformally invariant normal derivative~\cite{MR749522}. Crucially, the operator~$I\cdot D$ extends smoothly to conformal infinities encoded by the zero locus of the scale~$\sigma =0$. From the conical section picture of the Introduction and the interpretation of the Thomas-$D$ operator as the generalization of the ambient gradient operator,
it follows that the Laplace-Robin operator generates evolution along the~$\sigma$-direction, indeed this underlies standard Fefferman-Graham type expansions of the type crucial to the AdS/CFT correspondence~\cite{Gover:2011rz,Gover:2008sw}. Also, it is important to note that the weight of the tractor~$\Phi^\bullet$ will encode the mass of its underlying physical excitations~\cite{Gover:2008pt}, indeed the general mass Weyl-weight relationship for spin~$s$ fields is given by
\begin{equation}\label{massweight}
m^2=-\frac{2\J}{d}\, (w-s+2)(d+w+s-3)\, ,
\end{equation}
where for constant curvature spaces~$\frac{\scalebox{.7}{\J }\,\!}{d}=\frac\Lambda{(d-1)(d-2)}$. Massless fields appear when~$w=s-2$ while depth~$t$ PM ones arise at~$w=s-t-1$~\footnote{Note that the twist $\tau:=s-w$ and depth are related by $\tau=t+1$.}
(maximal depth~$t=s$ PM fields always have~$w=-1$).

Generally for higher spins, we are not interested in wave equations alone, but  must augment these with transversality conditions.
The first point to notice, is that as the spin increases, consistency of transversality requirements impose restrictions on the backgrounds
in which higher spin fields can propagate. We do not wish to  delve further into that issue here, so for the  remainder of this discussion concentrate on conformally flat spaces. This has the happy consequence that commutators of the Thomas-$D$ operator and scale tractor vanish~$[D^M,D^N]=0=[D^M,I^N]$ (the latter of these conditions of course holds more generally in conformally Einstein spaces).
Also, for massive spins, we desire a simple calculus automatically incorporating the St\"uckelberg fields required to describe them in  a gauge invariant way.
Let us sketch how this works for spins~1 and~2 before giving the equations we need at general~$s$ in Section~\ref{ths}.
For spin~1 we take as field content a weight~$w$ tractor~$A_M$ while for spin~2 we consider a weight~$w$ rank~2 symmetric tractor~$h_{MN}$ and
postulate gauge invariances mimicking their Maxwell and linearized general coordinate counterparts
$$
\delta A_M= D_M \alpha\, ,\qquad \delta h_{MN}=D_M \xi_N + D_N \xi_M\, .
$$
Because the Thomas-$D$ operator is null, under these transformations the ``Feynman- and Fock-de Donder-gauge'' parts of the fields~$A_M$ and~$h_{MN}$ are, respectively,
gauge inert, thus we may consistently impose conditions
\be
\label{deDonder}
D^{N}A_{N}= 0 \, , \qquad D^M h_{MN} -\frac 12 D_M h_N^N=0\, .
\ee
These conditions already ensure that the tractors~$A_M$ and~$h_{MN}$ are parameterized by: a vector and St\"uckelberg scalar for the Maxwell case; and metric fluctuations and  a
St\"uckelberg vector and scalar for the spin~2 case. For example, in the spin~2 case one finds gauge transformations for the metric fluctuations~$h_{\mu\nu}$ and St\"uckelberg fields~$(V_\mu,\varphi)$~\cite{Gover:2008pt,Gover:2008sw}
$$
\delta h_{\mu\nu}=\nabla_{(\mu} \xi_{\nu)}+\frac{2\scalebox{1}{\J}}d\, g_{\mu\nu} \,\xi\, ,\quad \delta V_\mu = w\,\xi_\mu + \partial_\mu \xi\, ,\quad \delta \varphi = (w+1)\,\xi\, .
$$
For generic weights (and in turn~$w$), the St\"uckelberg fields can be gauged away leaving a massive theory for~$h_{\mu\nu}$; when~$w=0$, the St\"uckelberg scalar~$\varphi$ can be gauged away and the vector~$V_\mu$ decouples leaving massless metric fluctuations~$h_{\mu\nu}$ with a linearized diffeomorphism gauge symmetry~$\delta h_{\mu\nu}=\nabla_{(\mu} \xi_{\nu)}$.
At~$w=-1$, the scalar decouples and the vector St\"uckelberg mode can be gauged away leaving residual symmetries with~$\xi_\mu=\partial_\mu \xi$. Under these, the metric fluctuations enjoy the PM gauge symmetry~$\delta h_{\mu\nu}= \big(\nabla_\mu\partial_\nu +\frac{2\scalebox{.7}{\J}}d\, g_{\mu\nu} \big)\xi$.

Oftentimes, a~$(d+1)$-dimensional projective approach based on a log-radial reduction~\cite{Biswas:2002nk} is employed to describe massive higher spins. In fact the above St\"uckelberg gauge transformations can be derived exactly in that way~\cite{Hallowell:2005np}. In the above description, the independent tractor field content is given by components~$h^{++}, h^{+m}$ and~$h^{mn}$. In fact quite generally,
the ``top slots'' of tractor fields encode the~$(d+1)$-dimensional projective construction~\cite{Gover:2008sw}. Geometrically this is easy to see; essentially one is projecting the Dirac cone along the scale tractor onto a surface of constant~$\sigma$. The images of conical sections at fixed values of~$\sigma$ are mapped in this way to loci with constant values of the log-radial
coordinate in a~$(d+1)$-dimensional hypersurface. These loci are again constant curvature manifolds.

The equations of motion for spins~1 and~2 are given by forming tractor analogs of the Maxwell curvature and Christoffel symbols
$$
F_{MN}:=D_M A_N - D_N A_M\, ,\qquad \Gamma^R_{MN}:=D_{(M}^{\phantom{R}}  h^R_{N)} - \frac12D^R h_{MN}\, ,
$$
and then coupling these to scale by simply contracting with the scale tractor
\be
\label{Fronsdal}
I^M F_{MN}=0\, ,\qquad I_R^{\phantom{R}} \Gamma^R_{MN}=0\, .\qquad
\ee
These equations of motion enjoy the above gauge invariances (so long as~$I_M \xi^M=0$ for spin~2) and have as leading terms  the universal Laplace-Robin structure~$I\cdot D\,  A_M + \cdots = 0$, ~$I\cdot D\,  h_{MN} + \cdots=0$. They encode massive, massless and PM equations  in a unified framework.

\section{Tractors and higher spins}\label{ths}

In this Section we apply   tractor technology to higher spin fields. In particular we show how to write wave equations and then
construct on-shell vertex functionals.

\subsection{On-shell higher spin tractors}\label{oshst}
 
For our ``on-shell'' purposes, the off-shell equations of motion presented for the special case of spin~$s=1,2$ cases
in the previous section are not optimal. Their on-shell counterparts are obtained by 
 fixing gauges for the St\"uckelberg auxiliaries:
 \[
  X^M \Phi_{MM_2\cdots M_s}=0\,,\qquad I^M \Phi_{MM_2\cdots M_s}=0\,,\qquad\Phi^M{}_{MM_3\cdots M_s}=0\,.
 \]
We then obtain the following equations, which generalize~\eqref{deDonder} and~\eqref{Fronsdal} directly to their higher~$s$ counterparts:
\be\label{eoms}
D^M \Phi_{MM_2\cdots M_s}=0\,,\qquad
I\cdot D\,  \Phi_{M_1\cdots M_s} = 0\,.
\ee
Here,~$\Phi_{M_1\ldots M_s}$ is a totally symmetric weight~$w$ tractor,
and masses and weights are related by~\eqref{massweight} above.
For tuned weights
$$
w=-1,0,\ldots,s-t-1,\ldots,s-2\, ,
$$
the above  on-shell equations describe
depth~$t$ PM and massless~$(t=1)$ excitations. At these weights,  residual gauge invariances appear~\cite{Gover:2008pt,Gover:2008sw}
\begin{equation}\label{gauges}
\delta\,\Phi_{M_1\cdots M_s}=D_{(M_1}\cdots D_{M_t}\, \Xi_{M_{t+1}\cdots M_s)}\, ,
\end{equation}
where the gauge parameters~$\Xi$ obey exactly the same set of conditions as the fields~$\Phi$ listed in~\eqref{eoms}.
The on-shell equations of motion~\eqref{eoms} and their residual invariances~\eqref{gauges} give the description of spin~$s$ fields
needed for our vertex calculations. 

Our next step is a simple technical man\oe uvre. Totally symmetric tensors~$\varphi_{\mu_1\cdots \mu_s}$ written in 
a ``symmetric-form'' notation~$\varphi(x,dx):=\varphi_{\mu_1\cdots \mu_s}(x)\, dx^{\mu_1\cdots \mu_s}$ can be treated
as functions of coordinates~$x^\mu$ and commuting differentials~$dx^{\mu}$. By introducing also derivatives with respect
to the differentials
$\partial/\partial(dx^{\mu})$, the main operations on symmetric tensors (the symmetrized gradient, divergence and trace) 
can be handled in an efficient, index free way~\cite{Damour:1987vm,Vasiliev:1988xc,Hallowell:2005np,Hallowell:2007zb}. This also allows physical quantities  to be described as generating functions 
simultaneously describing all
spins~$s$. The same methods can be applied in the ambient space (or tractor bundle fibre)~\cite{Grigoriev:2011gp} by 
 re-expressing symmetric tractor fields~$\Phi_{M_{1}\cdots M_{s}}$ as
~$$
	\Phi(x,U):=\,\Phi_{M_{1}\cdots M_{s}}(x)\,U^{M_{1}}
	\,\cdots\,U^{M_{s}}\,.
$$
In these terms, the tracefree condition takes the form 
$\partial_{U}\cdot \partial_U\,\Phi(x,U)=0$\,.
Also, we will need the operator whose eigenvalue is the spin~$s$ of~$\Phi(x,U)$, this is 
simply~$U\cdot \partial_U$. Finally, the difference of the spin and weight appears in
many places, therefore we define the twist\footnote{Note that the twist~$\tau$ is related to the homogeneity~$\m$ of~\cite{Joung:2012rv,Joung:2012fv,Taronna:2012gb,Joung:2012hz}  by~$\t=\m+2$.}
$$
\tau:=s-w\, .
$$
We have summarized the higher spin field equations in index-free tractor notation in Figure~\ref{tfree}.

\begin{figure}
\begin{center}
\shabox{
\begin{tabular}{lcc}
$I\cdot \wD\,\Phi(x,U)= 0\,,$ &
	$\partial_{U}\cdot \partial_{U}\, \Phi(x,U)=0\,,$ &~$U\cdot \partial_U\, \Phi(x,U)=\, s\, \Phi(x,U)\, ,$\\[3mm]
	$D\cdot\partial_{U}\, \Phi(x,U)=0\, ,$ &~$X\cdot\partial_{U}\,\Phi(x,U)=0\, ,$ &~$ I\cdot\partial_{U}\,\Phi(x,U)=0\, .$
\end{tabular}
}
\end{center}
\caption{\label{tfree} Index-free tractor field equations for totally symmetric higher spins of any mass type.
These also apply to the ambient description in terms of fields~$\Phi(X,U)$ extended off the cone and subject to 
$
\Phi\sim \Phi+X^2 S\, . 
$ In that case the weight condition is rewritten as the homogeneity one~\eqref{hom}.}
\end{figure}

\subsection{Conformally invariant functionals}
\label{abcd}

We want to establish a basis for the possible on-shell vertex functions. These can be written efficiently 
by splitting the spacetime points associated with each on-shell field.
Point-split, on-shell, densities can then be written in terms of tractors in much the same way as is
done for standard tensors. The key differential operator is  the Thomas-$D$.
Indeed, any quantity which is of the~$N$-th order in on-shell tractors
and involves~$\d$ Thomas-$D$ operators can be expressed in the form
\ba\label{den L}
	\mathcal L[\Phi_{1},\cdots,\Phi_{N}](x) \eq
	\prod_{i=0}^{\frac{N-\delta}2}\prod_{j=0}^{\delta}\,
	 \partial_{U_{n_{i}}}\!\!\cdot\partial_{U_{m_{i}}}\,
	 \partial_{U_{n_{j}}}\!\!\cdot \wD_{m_{j}}\times\nn
	 &&\qquad \times\ 
	\Phi_{1}(x_{1},U_{1})\,\cdots\,\Phi_{N}(x_{N},U_{N})\,
	\Big|_{\overset{X_{1}=\cdots X_{N}=X}{\sst U_{1}=\cdots=U_{N}=0}}\, .
\ea
The conformal weight (or degree of homogeneity) of the above density is~$w_{1}+\cdots +w_{N}-\d$,
where each~$w_{i}$ labels the weight of the tractor field~$\Phi_{i}$\,. Subscripts~$i,j,\ldots$  will generally be used to label points in the point-splitting procedure.

Generically, to establish a complete dictionary between (pseudo-)Riemannian and tractor quantities, one also employs the canonical tractor~$X^M$. 
However, point-split quantities involving the canonical tractor can be recast in terms of the above basis of invariants.
This vastly simplifies our ansatz for cubic interactions, so let us elaborate on this point:
One could consider operators ~$X\cdot \wD_{\sst i}$ or~$X\cdot \partial_{U_{i}}$ acting on~\eqref{den L}.
But, because  this expression is to be evaluated at a single point~\mt{x_i=x}\,, 
these operators can be traded for~$X_{i}\cdot \wD_{\sst i}$ or~$X_{i}\cdot \partial_{U_{i}}$\,, the operators entering the homogeneity~\eqref{hom} and tangentiality conditions (see the second line of the display in Figure~\ref{tfree}) respectively. The former then just returns the degree of homogeneity while the latter annihilates the tensor field~$\Phi(x_{i},U_{i})$ on-shell.  Therefore, to eliminate~$X_{i}\cdot\partial_{U_{i}}$'s it suffices to commute them with all Thomas-$D$ operators acting on~$\Phi(x_{i},U_{i})$\,, 
which can be achieved via the commutator/reordering identity displayed in Eq.~\eqref{XDcomm}.
In our index free notation, this implies
\be\label{x cmr}\nonumber
	\big[X\cdot\partial_{U}\, , \,  \wD_{\sst M}
	\big]=\frac2{h}\,X_{\sst M} \,\partial_U\cdot  \wD-\partial_{U^{M}}\, ,
\ee
which generates no new~$X$-dependence on-shell, the latter being proportional to a divergence operator.

To construct vertices, we still need to integrate densities such as~\eqref{den L} over slices of the cone,
in a way that maintains manifest Weyl invariance (of course this is ultimately broken as explained earlier
by the coupling to scale~$\sigma$). For that we observe that the
$d$-dimensional measure~$\sqrt{-g}\,\sigma^{-d}$ is Weyl invariant, so that for any
Weyl invariant function~$f$, the integral~$F[g,\sigma]=\int d^{d}x \sqrt{-g}\,\sigma^{-d}\,f(g,\sigma)=F[\Omega^2\, g,\Omega\, \sigma]$
is also Weyl invariant.\footnote{Then, the parallel scale tractor construction ensures that choosing~$\Omega$ such that ~$\sigma=1$
 singles out the underlying Einstein metric from the conformal class~$[g]$. }
We will denote 
 \begin{equation}\label{measure}
 \int_\sigma f:=\int_{\cM}\!d^{d}x\, \frac{\sqrt{-g}}{\sigma^d}\,  f\,.
\end{equation}
Thus, to integrate the quantity~\eqref{den L}, we must first convert it to the correct weight by introducing 
 a suitable power of~$\sigma$:
\be\nonumber
	\int_{\sigma}\,
	\sigma^{\d-w_{1}-\,\cdots\,-w_{N}}\,\mathcal L[\Phi_{1},\ldots,\Phi_{N}]\,.
\ee
This  functional can be used for the construction of both actions and vertices.
For, example, the leading quadratic term of the on-shell action is captured by
$$
	S^{\sst (2)} \propto \int_{\sigma}\,\sigma^{1+2\,\mu}\,
	e^{\s^{-2}\,\partial_{U_{1}}\!\cdot\,\partial_{U_{2}}}\,
	\Phi(x,U_{1})\ I\cdot \wD\ \Phi(x,U_{2})\,\Big|_{U_{1}=U_{2}=0}\, .
$$
It is not difficult to show by employing the harmonic gauge of Appendix~\ref{harmgauge} that this expression is equivalent to the 
one obtained in a~$(d+1)$-dimensional projective formulation in~\cite{Joung:2011ww}.

Finally, the tractor integration by parts formula~\eqref{parts} 
implies the following integration by parts rule for our vertex functionals:
\be\label{int-b-p}
	(h_B-2)\,\int_{\sigma}\,\s^d\, A\,(\wD_{\sst M}\,B^{\sst M})
	=(h_A-2)\,\int_{\sigma}\,\s^d\,(\wD_{\sst M}\, A)\, B^{\sst M}\,,
\ee
where the weights~$h_{A}$ and~$h_{B}$ of~$A$ and~$B^{\sst M}$ satisfy the relation \mt{h_{A}+h_{B}=2}\,. This property together with the deformed Leibniz rule~\eqref{Leibniz} and the commutation relation~\eqref{x cmr} constitute a complete vertex calculus.

\section{Cubic interactions}\label{Cubic}

In order to construct  interactions of higher spin fields, we rely on gauge invariance.
This is based on the assumption that a non-linear deformation of the leading gauge symmetries
is responsible for propagation of the correct higher spin physical degrees of freedom.
Gauge invariance of the interacting theory can be analysed  perturbatively by expanding the (ultimate non-linear) action in powers of the gauge fields following a Noether-type procedure.

\subsection{Noether procedure}

In the standard setup, one considers an expansion of the gauge invariant action and its non-linear gauge symmetry order by order in the number of gauge fields.
Up to the cubic order, these are
\be\label{act exp}\nonumber
	S=S^{\sst (2)}+S^{\sst (3)}+\cdots\,,\qquad
	\delta\,\Phi=\delta^{\sst (0)}\,\Phi + \delta^{\sst (1)}\,\Phi+
	\cdots\,.
\ee
Assuming that~$\delta S=0$, then
it follows that
\be\label{nt eq}
	\delta^{\sst (0)}\,S^{\sst (n)}+
	\delta^{\sst (1)}\,S^{\sst (n-1)}
	+\cdots +\delta^{\sst (n-2)}\,S^{\sst (2)}=0\,, \qquad [n\ge2]\,.
\ee
Our ultimate aim is to solve these conditions iteratively.
Focusing on the first non-trivial part of~\eqref{nt eq} gives the requirement relevant for our current cubic problem:
\be\nonumber
	\delta^{\sst (0)}\,S^{\sst (3)}+\delta^{\sst (1)}\,S^{\sst (2)}=0\,.
\ee
This task is further simplified by observing that linearly on-shell (denoted~$\approx$)
\mt{\delta^{\sst (1)}\,S^{\sst (2)}\approx 0}\,. 
Therefore, the first step of the Noether procedure is to solve
\be\label{nt cub}
	\delta^{\sst (0)}\,S^{\sst (3)}\approx 0\, .
\ee
This problem enjoys an elegant and simple tractor-based solution.

\subsection{The cubic vertex ansatz}

To solve the cubic-order gauge consistency condition~\eqref{nt cub}, we start with the most general transverse and traceless, parity-invariant,~$d$-dimensioanl cubic interactions~$S^{\sst (3)}$. In generating function notation, these take the compact form
\be\label{gen cub}
	S^{\sst (3)}=
	\int_{\sigma}\,\colon\,
	C(\,\sigma\,,\,\partial_{U_{i}}\,,\,\wD_{\sst i}\,)\,\colon\ 
	\Phi(x_{1},U_{1})\,\Phi(x_{2},U_{2})\,\Phi(x_{3},U_{3})\,
	\Big|_{\overset{x_i=x}{\sst U_{i}=0}}\,,
\ee
where we neglect terms involving St\"uckelberg or auxiliary fields. 
Here~$\colon \,C\,\colon$ denotes a normal ordered operator.
The normal ordering can be chosen such that~$\sigma$ sits to the left. Although, passing the Thomas-$D$ operator through
powers of~$\sigma$ can produce the scale tractor~\mt{I^{\sst M}=\wD^{\sst M}\sigma}, the dependence on it can be removed 
by noticing that it can appear in combinations removable by linear order field equations:
\begin{equation*}
	I\cdot\partial_{U_{i}}\approx  0\,, \qquad
	I\cdot \wD_{i}\approx 0\,.
\end{equation*}
The remaining~$6(d+2)$ variables appearing in the operator~$C$ can be packaged in twelve combinations:
 \begin{equation*}
	\colon \,C(\,\sigma\,,\,\partial_{U_{i}}\,,\,\wD_{\sst i}\,)\,\colon\
	\approx \ \colon\, C(\,\sigma\,,\,\X\,,\,
	\Y\,,\,
	\Z\,)\,\colon\,.
\end{equation*}
Here, the operators~$\X,\Y,\Z$ are point-split combinations of internal index and Thomas-$D$ operators given by
\[
\X_{ij}=\wD_i \cdot\wD_j\,,\qquad
\Y_{ij}=\partial_{U_i}\cdot\wD_j\, ,\qquad
\Z_{ij}=\partial_{U_i}\cdot\partial_{U_j}\,,
\]
and  they satisfy~$\X_{ij}=\X_{ji}$\,,~$\Z_{ij}=\Z_{ji}$ and
$\X_{ii}\,\Phi_{i}\approx 0$,~$\Y_{ii}\,\Phi_{i}\approx 0$,~$\Z_{ii}\,\Phi_{i}\approx 0$\,.
We choose the remaining normal orderings according to~$\sigma>\X>\Y>\Z>h$.
The twelve variables~$(\X,\Y,\Z)$ can be halved because all three~$\X_{ij}$'s as well as the half of the~$\Y_{ij}$'s, say the~$\Y_{i\,i-1}$'s, can be removed by re-expressing them in terms of the others. To prove this requires a set of identities that we develop in Appendix~\ref{reorder}. Thus our vertex ansatz now reads
\begin{equation}\label{simple}
	S^{\sst (3)}\approx 
	\int_{\sigma}\,\colon\,
	C(\,\sigma\,,\,\Y_{i}\,,\,\Z_{i}\,)\,\colon\ 
	\Phi(x_{1},U_{1})\,\Phi(x_{2},U_{2})\,\Phi(x_{3},U_{3})\,
	\Big|_{\overset{x_i=x}{\sst U_{i}=0}}
	=:
	\big\langle \,C(\,\sigma,\,\Y_i\,,\,\Z_i\,)\,\big\rangle_{\Phi_{1}\Phi_{2}\Phi_{3}}\,,
\end{equation}
where
\be\label{var}\nonumber
\Z_i\,=\,\s^{-2}\,\Z_{i-1,i+1}\,,\qquad \Y_{i}\,=\,\Y_{i,i+1}\,,\qquad[i\sim i+3]\,.
\ee

Having removed all redundant variables we can now determine the~$\s$-dependence of the vertex by simply counting the homogeneity degree of generic monomials
\begin{equation*}
\Y_1^{s_1-m_2-m_3}\,\Y_2^{s_2-m_3-m_1}\,\Y_3^{s_3-m_1-m_2}\,\Z_1^{m_1}\,\Z_2^{m_2}\,\Z_3^{m_3}\,.
\end{equation*}
Since
$
(X\cdot\wD-U\cdot\partial_U+\tau)\,\Phi\,=\,0
$,
 the above monomial has homogeneity
$-(\tau_{1}+\tau_{2}+\tau_{3})$. 
Therefore, it follows that the vertex dependence of~$\sigma$ is simply
\begin{equation}\label{cubint}
S^{(3)}\,\approx \,\big\langle\,\s^{\tau_{1}+\tau_{2}+\tau_{3}}\,
C(\,\Y_i\,,\,\Z_i\,)\,\big\rangle_{\Phi_{1}\Phi_{2}\Phi_{3}}\, .
\end{equation}

\subsection{Vertex gauge invariance}\label{vgi}

For vertices where all external lines are massive and on-shell, we can go no further with our on-shell, three-point, analysis of allowed higher spin vertices. The reason is that, unlike their massless and PM counterparts, which enjoy a residual, onshell,  gauge invariance~\eqref{gauges}, 
the massive fields only obey second class constraints which have already been implemented at this order. (Of course, one could either study 
off-shell three point vertices or higher point functions, to further whittle down the space of cubic vertices concordant with massive constraints.)
Hence we now focus on the case where at least one external line has a residual gauge invariance.

Let us first focus on the case when one field is massless:~$\tau_1=2$ (say), and try to  solve~\eqref{nt cub} by requiring gauge invariance. To perform a linear gauge variation of the cubic interaction~\eqref{cubint} with respect to the field~$\Phi_1$ we replace
$$\Phi_1\ \longrightarrow\ \delta^{\sst (0)}_{E_{1}}\Phi_{1}=U_1\cdot \wD_1\,E_1\, ,$$ by its (strictly massless) gauge 
transform corresponding to Eq.~\eqref{gauges} at $t=1=\tau-1$. Then we must push the operator $U_{1}\cdot \wD_{1}$ to the left where it 
vanishes because the integrand is evaluated at $U_{1}=0$\,. 
For this we employ the
 commutation relations\footnote{We display only  non-vanishing commutators; these can be easily computed using the operator algebra generated by $U_i$ and $\partial_{U_j}$.}:
\begin{equation*}
[\,\Y_i\,,\,U_i\cdot \wD_i\,]\,=\,\X_{i,i+1}\,,\qquad 
[\,\Z_{i\pm1}\,,\,U_i\cdot \wD_i\,]\,=\,\s^{-2}\,\Y_{i\mp1,i}\,,
\end{equation*}
in order to encode this man\oe vre in terms of ordinary derivatives on the function $C$ that labels the ``vertex operator''.
Orchestrating these manipulations we get:
\begin{equation*}
\delta^{(0)}_{E_1}\,S^{(3)}\,\approx \,\left\langle\, \s^{2+\tau_2+\tau_{3}}\left[\X_{12}\,\partial_{\Y_1}+\s^{-2}\,\Y_{31}\,\partial_{\Z_2}+\s^{-2}\,\Y_{21}\,\partial_{\Z_3}\right]\,C(\,\Y_i\,,\,\Z_i\,)\,\right\rangle_{E_{1}\Phi_{2}\Phi_{3}}\,.
\end{equation*}
Then, by applying the identities~\eqref{chi id} and~\eqref{total deriv}, with~$\a-1$ and~$\beta+1$ given by the operator
\begin{equation*}
\hat \g:=\tau_{2}+\tau_{3}-d+1-2\,\sum_{i}\Z_i\,\partial_{\Z_i}\,,
\end{equation*}
we can rewrite this variation in terms of the restricted set of variables $\Y_i$ and $\Z_i$ as
\begin{multline*}
\delta^{(0)}_{E_1}\,S^{(3)}\,\approx 
\left\langle \s^{\tau_{2}+\tau_{3}}\Big[\Y_3\partial_{\Z_2}-\Y_2\partial_{\Z_3}\right.\\ \left.-\,\hat\g \left(\Y_3\partial_{\Y_3}-\Y_2\partial_{\Y_2}+\tfrac{\tau_2-\tau_3}2\right)\partial_{\Y_1}\Big]\,C(\,\Y_i\,,\,\Z_i\,)\right\rangle_{E_{1}\Phi_{2}\Phi_{3}}\,=\,0\, .
\end{multline*}
From this we  extract the differential equation
\begin{equation*}
\Big[\Y_3\partial_{\Z_2}-\Y_2\partial_{\Z_3} \,-\, \hat\g\left(\Y_3\partial_{\Y_3}-\Y_2\partial_{\Y_2}+\tfrac{\tau_2-\tau_3}2\right)\partial_{\Y_1}\Big]\,C(\,\Y_i\,,\,\Z_i\,)\,=\,0\, ,
\end{equation*}
which can be slightly rewritten using 
\begin{equation*}
\Big\langle \s^{\tau_{2}+\tau_{3}}\ \hat\g\ \partial_{\Y_1}C(\,\Y_i\,,\,\Z_i\,)\Big\rangle_{E_{1}\Phi_{2}\Phi_{3}}
=\,-\,\left\langle\ \hat \d\ \s^{\tau_{2}+\tau_{3}}\,\partial_{\Y_1}C(\,\Y_i\,,\,\Z_i\,)\right\rangle_{E_{1}\Phi_{2}\Phi_{3}}\,,
\end{equation*}
where
\begin{equation}\label{dhat}
\hat\delta\,:=\,-\,\s\,\frac{d}{d\s}\,+\,d-1\,
=-\,\s^{d}\,\frac{d}{d\s}\,\s^{-d+1}\, .
\end{equation}
This gives our formula of the linear gauge variation of the cubic vertex:
\begin{multline*}
\delta^{(0)}_{E_1}\,S^{(3)}\,\approx \,\left\langle \Big[\Y_3\partial_{\Z_2}-\Y_2\partial_{\Z_3}\,+\,\hat\d \left(\Y_3\partial_{\Y_3}-\Y_2\partial_{\Y_2}+\tfrac{\tau_2-\tau_3}2\right)\partial_{\Y_1}\Big]\times\right.\\\left.\times\,
\s^{2+\tau_{2}+\tau_{3}}\,C(\,\Y_i\,,\,\Z_i\,)\right\rangle_{E_{1}\Phi_{2}\Phi_{3}}\,=\,0\, .
\end{multline*}
All in all, this gives our final result for the differential equation determining cubic interactions: 
\be
\Big[\Y_3\partial_{\Z_2}-\Y_2\partial_{\Z_3}\,+\,\hat\d \left(\Y_3\partial_{\Y_3}-\Y_2\partial_{\Y_2}+\tfrac{\tau_2-\tau_3}2\right)\partial_{\Y_1}\Big]\,C(\,\Y_i\,,\,\Z_i\,)\,=\,0\,,\label{tractor cons}
\ee
where~$\hat\d$ can be considered here as an auxiliary variable on which the function~$C(\,\Y_i\,,\,\Z_i\,)$ depends. It can be substituted for its operator definition~\eqref{dhat} at the last step.
The above equation coincides with the consistency condition obtained using~$(d+1)$-dimensional projective methods~\cite{Joung:2012fv}.

The above discussion is quite general and extends also to PM couplings along the same lines as in~\cite{Joung:2012fv}. Indeed, since the gauge transformations of the PM fields are multiple gradients with respect to the Thomas-$D$ operator (see Eq.~\eqref{gauges}), at the PM point~$\tau_1\in\mathbb{N}$ the corresponding differential equation factorizes  as:
\be\nonumber
\prod_{n=0}^{\tau_1-2}\Big[\Y_3\partial_{\Z_2}-\Y_2\partial_{\Z_3}\,+\,\hat\d \left(\Y_3\partial_{\Y_3}-\Y_2\partial_{\Y_2}+\tfrac{\tau_1+\tau_2-\tau_3-2n-2}2\right)\partial_{\Y_1}\Big]\,C(\,\Y_i\,,\,\Z_i\,)\,=\,0\, .\label{tractor cons PM}
\ee

\section{Conclusions}

In this Article we have addressed the problem of constructing higher-spin cubic interactions for totally symmetric fields using tractor calculus. This extends and generalizes the results obtained requiring (Stueckelberg-)gauge invariance in~\cite{Joung:2011ww,Joung:2012rv,Joung:2012fv,Taronna:2012gb} and~\cite{Joung:2012hz}. (The latter PM analysis completes the flat space, light-cone, cubic interaction program of~\cite{Metsaev:1993ap,Fradkin:1995xy,Metsaev:2005ar,Metsaev:2007rn}.) 
Asides from deeper questions involving the higher dimensional nature of spacetime and the {\it r\^ole} of conformal geometry, it also clarifies and simplifies the~$\d$-function radial integrations in the aforementioned projective space approaches. In particular, 
the projective measure factor~$\d(\sqrt{X^2}-L)$ is replaced by a standard~$d$-dimensional one. The projective space tensor structures  are 
then encoded by sections of the~$(d+2)$-dimensional tractor bundle. Moreover, non-conformal interacting higher spins now couple in a conformally covariant way to scale through the scale tractor~$I$, the very tensor encoding the underlying geometry.

The end result can be summarized by a simple and complete dictionary between the projective-space, cubic vertex function~$C(Y,Z)$ and measure (of~\cite{Joung:2011ww,Joung:2012rv,Joung:2012fv,Taronna:2012gb,Joung:2012hz}) and their tractor counterparts given by
\ba
\int d^{d+1}X\ \delta(\sqrt{X^2}-L)\qquad&\longleftrightarrow&\qquad 
\int_{\s}:=\int d^{d}x\,\frac{\sqrt{-g}}{\s^d}\,,\nn[2mm]
Y_i=\partial_{U_i}\cdot\partial_{X_{i+1}} \qquad&\longleftrightarrow&\qquad\Y_i=\partial_{U_i}\cdot\wD_{i+1}\,,\nn[3mm]
Z_i=\partial_{U_{i-1}}\!\!\cdot\partial_{U_{i+1}}\qquad&\longleftrightarrow&\qquad \Z_i=\s^{-2}\ \partial_{U_{i-1}}\!\!\cdot\partial_{U_{i+1}}\,.\nonumber
\ea
This result likely carries over to higher point functions and therefore provides a useful avenue to extend the higher point analysis of~\cite{Taronna:2011kt,Taronna:2012gb}. It also hints at a (dual) $(d+2)$-dimensional field theory~\cite{Gover:2009vc} underlying higher spin interactions. Let us stress here that cubic consistency alone does not  control the second class constraints of massive or PM fields. In  both massless and massive cases, quartic consistency is  expected to further restrict cubic couplings  (see, {\it e.g.},~\cite{Boulanger:2013zza}).

A distinct advantage of the tractor approach is that it makes both bulk and boundary conformal structures explicit.
Since, our analysis of bulk vertices  will ultimately be dual to the boundary conformal block-type analyses of, for example,~\cite{Ferrara:1972uq,Dolan:2000ut,Costa:2011dw,SimmonsDuffin:2012uy,Stanev:2012nq,Zhiboedov:2012bm}, this indicates the existence 
of a new dictionary between boundary correlators and bulk Witten diagrams realized as different gauge fixings of the same tractor expression. It is also interesting to point out the simpler nature of bulk inputs with respect to their CFT counterparts that solve more complicated differential equations~\cite{Zhiboedov:2012bm}. In fact, the ambient construction of~\cite{Bekaert:2012vt,Bekaert:2013zya} is a first step in this direction. Thus, a detailed analysis would start by clarifying the relations between our bulk results and the bulk normalizable and non-normalizable solutions and the corresponding CFT operators and shadow fields given there. Also, the solution generating algebra of~\cite{Gover:2011rz,RodGover:2012ib} gives simple formul\ae\  for the bulk boundary propagators for those fields. In fact, even though both in the CFT side and in the bulk side one is able to classify the corresponding current correlators and bulk couplings respectively, a precise dictionary relating the two is still unavailable as are the corresponding Witten diagrams (see however \cite{Colombo:2010fu,Colombo:2012jx,Didenko:2012tv,Gelfond:2013xt,Didenko:2013bj} for interesting examples of n-point correlation function computations exploiting higher spin symmetry). This construction is important to clarify the structure of cubic couplings in Vasiliev's system as well in the more complicated case of String Theory. Another issue is locality beyond quartic order (where ever increasing towers of derivative interactions can set in). Also a bulk understanding of the conformal bootstrap 
(which determines CFT correlators from cubic vertices, associativity and conformal invariance) would be desirable~\cite{Taronna:2012gb}.
Because it connects bulk and boundary ambient approaches through their conformal structure, the tractor approach can indeed cast 
light on all these issues.

\section*{Acknowledgements}

E.J. and A.W. thank the Schr\"odinger  Institute, Vienna, for the hospitality during the ``Workshop on Higher-Spin Gravity''. A.W. acknowledges the Scuola Normale Superiore, Pisa for hospitality during the ``Workshop on Supersymmetry, Quantum Gravity and Gauge Fields''. M.T. is grateful to Itzhak Bars for discussions. All the authors thank the GGI, Florence workshop on ``Higher spins, Strings and Duality'' where this Article was completed. The work of E.J. was supported in part by Scuola Normale Superiore, by INFN, and by the MIUR-PRIN contract 2009-KHZKRX.

\appendix

\section{Reordering identities}\label{reorder}

In this Appendix, we provide the operator identities required to reach the simple ansatz for the 
cubic vertex given in Eq.~\eqref{cubint}.

\subsubsection*{The~$\X_{ij}$ operators}

In the following we are going to show that any~$\X_{ij}$ can be removed in terms of the other operators. Without loss of generality, consider~$\X_{12}$:
\begin{equation*}
\big\langle\,\s^{d+\a}\,\X_{12}\,F(\X,\Y,\Z)\,\big\rangle
=:\mathscr A\,,
\end{equation*}
where 
the power~$\a$ is determined by requiring
integrands to have the correct weight.
The generalized Leibniz rule~\eqref{Leibniz} together with the on-shell condition on the fields gives
\[
\wD_1\cdot \wD_2\approx-\tfrac{1}{2}\,\sigma^{-1}\,I\cdot D_{12}\,,
\]
where the subscript~$ij$ of~$\wD_{ij}$ means that  
it acts as~$\wD(\Phi_{i}\Phi_{j})$\,.
Using this identity, one gets
\begin{equation*}
\mathscr A
\approx\,-\tfrac{1}{2}\,
\big\langle\,\s^{d+\a-1}\,I\cdot D_{12}\,F(\X,\Y,\Z)\,\big\rangle\,.
\end{equation*}
Then, using the integration by parts formula~\eqref{int-b-p}, this becomes
\begin{equation*}
\mathscr A\approx\,-\tfrac{1}{2}\,
\big\langle\,\s^{d}\,I\cdot D_{3}\,\s_{3}^{\a-1}\,F(\X,\Y,\Z)\,\big\rangle\,.
\end{equation*}
Upon using the identity, valid for~$f\in {\rm ker} (I\cdot\wD)$,
\begin{equation}\label{Ds}
[\wD^M\,,\,\sigma^k]\,f\,=\,k\,\sigma^{k-2}\,\left(\sigma\,I^M-\tfrac{k-1}{h}\,I^2\,X^M\right)\,f\,,
\end{equation}
we finally find
\be\label{chi id}
\big\langle\,\s^{d+\a}\,\X_{12}\,F(\X,\Y,\Z)\,\big\rangle
\approx\,-\tfrac{1}{2}\,(\a-1)\,
\big\langle\,\s^{d+\a-2}\,(h_{3}+\a-2)\,F(\X,\Y,\Z)\,\big\rangle\,,
\ee
which allows us to  eliminate completely the dependence on~$\X_{12}$.

\subsubsection*{The~$\Y_{ij}$ operators}

We want to  show that any~$\Y_{i,i+1}$ can be replaced by combinations of ~$\Y_{i,i-1}$ and the other operators.
Focusing on~$\Y_{12}$, consider
\begin{equation*}
\big\langle\,\s^{d+\b}\,\Y_{12}\,F(\X,\Y,\Z)\,\big\rangle
=:\mathscr B\,.
\end{equation*}
Using the identity~\eqref{Ds}
we find
\begin{equation*}
\mathscr B
=
\left\langle \s^{d}\left(\partial_{U_{1}}\!\cdot D_{2}\,\s_{2}^{\b}
+\b(\b-1)\,\s^{\b-2}\,X\cdot\partial_{U_{1}}\right)\tfrac1{h_{2}+2\b-2}\,F(\X,\Y,\Z)\,
\right\rangle.
\end{equation*}
Integrating by parts this becomes
\begin{equation*}
\mathscr B
=\left\langle\s^{d+\b}\left(\partial_{U_{1}}\!\cdot D_{13}
+\b(\b-1)\,\s^{-2}\,X\cdot\partial_{U_{1}}\right)\tfrac1{h_{2}+2\b-2}\,F(\X,\Y,\Z)\,
\right\rangle.
\end{equation*}
This time the generalized Leibniz rule~\eqref{Leibniz} gives
\begin{equation*}
\partial_{U_1}\!\cdot \wD_{13}\,=\,\partial_{U_1}\!\cdot \wD_{1}+
\partial_{U_1}\!\cdot \wD_{3}-\tfrac2{h_{13}}\,X\cdot\partial_{U_1}\,\wD_1\cdot\wD_3\,,
\end{equation*}
and hence
\[
\mathscr B
\approx \left\langle\s^{d+\b}\left[h_{13}\,\partial_{U_{1}}\!\cdot \wD_{3}
-\left(2\,\wD_{1}\cdot\wD_{3}-\b(\b-1)\,\s^{-2}\right)X\cdot\partial_{U_{1}}\right]\tfrac1{h_{2}+2\b-2}\,F(\X,\Y,\Z)\,
\right\rangle.
\]
Here, we can trade~$X\cdot\partial_{U_1}$ for~$X_1\cdot\partial_{U_1}$ at the price of a commutator.
Using~\eqref{XDcomm}
together with the identity~\eqref{chi id} for~$\wD_1\cdot\wD_3$ one ends up with
\begin{equation*}
\mathscr B\approx 
\left\langle \s^{d+\b}\left(\tfrac{h_{13}+2}{h_{2}+2\b-2}\,\Y_{13}
+(\b-1)\s^{-2}\,X_{1}\cdot\partial_{U_{1}}\right) F(\X,\Y,\Z)\,
\right\rangle.
\end{equation*}
Finally, using the fact that the overall vertex has zero weight
\begin{equation*}
d+\b+\tfrac12\,(h_2-d)+\tfrac12\,(h_{13}-d)=0\,,
\end{equation*}
and pushing to the right~$X_1\cdot\partial_{U_1}$ one gets
\[
\big\langle \,\s^{d+\b}\,(\Y_{12}+\Y_{13})\,F(\X,\Y,\Z)\,\big\rangle
\approx \,(1-\b)\,\big\langle\,\sigma^{d+\b-2}
\left(\Z_{i1}\partial_{\Y_{i1}}+\Y_{1i}\partial_{\X_{1i}}\right)\,F(\X,\Y,\Z)\,
\big\rangle\,,
\]
that can be turned into a formula for total derivatives as
\be\label{total deriv}
 \left\langle\s^d\left[\,\partial_{U_1} \cdot (\wD_1+\wD_{2}+\wD_3)\,+\,\frac{\partial}{\partial \s}\,\frac{1}{\s}\,\Z_{2}\partial_{\Y_{3}}\,\right]\,
 \s^\b\,F(\Y_{i},\Z_{i})\right\rangle\approx \,0\,.
\ee

\section{Ambient harmonic gauge}\label{harmgauge}

To connect our tractor approach with previous~$(d+1)$-dimensional projective space methods,  we 
work in the Fefferman-Graham ambient space described in Section~\ref{ambient} and employ a harmonic gauge choice for how tractors are extended off the cone. 
In this Appendix, we review the ambient description of higher spin wave equations and then show how the harmonic gauge connects
our vertex result with cubic interactions known via projective space methods.

\subsection{Ambient wave equations}

Our basic objects are now ambient space tractor generating functions~$\Phi(X,U)$ subject to the equivalence relation~\eqref{cone}. Representatives of these equivalence classes can be chosen by fixing a gauge, an enlightening choice being
the ambient harmonic condition
$$
\partial_{M}\partial^M \,\Phi(X,U)\,=\,0\, .
$$
As a welcome consequence of~\eqref{tom D}, in this gauge the ambient Thomas-$D$ operator can be replaced by the gradient 
$$
\wD_M\rightarrow \partial_M\, .
$$
The higher spin equations of motion~\eqref{eoms} then have simple interpretations: The Laplace--Robin condition~$I\cdot D\ \Phi=0$
says
$$
I\cdot \partial_X\,  \Phi(X,U) = 0\, .
$$
In the conformally-flat setting~$I^{\sst M}$ is a constant vector so the 
 ambient space dependence of~$\Phi$ is reduced to the~$(d+1)$-dimensional hyperplane~$\mathcal P^{d+1}$\,:
\begin{equation*}
	\mathcal P^{d+1}:=\{ X\in\widetilde \cM\,|\ \sigma(X)=\mbox{constant}\,\}\,,
	\qquad \sigma(X):=I\cdot X\, .
\end{equation*}
Typically we choose~$\sigma=1$, the crucial point is that the choice~$\sigma=0$ should be avoided
as it corresponds to the boundary in an AdS setting and is singular for dS and Minkowski spaces---physically this corresponds to the 
choice of units for the Planck constant.
Then, depending on the direction (time-like, space-like or light-like) of
the ambient scale tractor~$I^{\sst M}$ the intersection between~$\mathcal P^{d+1}$ and the cone~$\cC^{d+1}:=\{ X\in \widetilde \cM\,|\,X^2=0\}$\,:
\begin{equation*}
	\cM:=\mathcal P^{d+1}\cap \cC^{d+1}\,,
\end{equation*}
gives all the maximally symmetric spaces in~$d$-dimensions (respectively dS, AdS or Mink\-owski).
This establishes that our tractor description amounts to fields living in constant curvature spaces. 
The tangentiality conditions~$X\cdot\partial_U\ \Phi = 0 = I\cdot \partial_U\ \Phi$ then reduce the tractor tensor 
multiplets to standard tensor ones and the tractor trace~$\partial_U\cdot\partial_U \ \Phi= 0$, in turn, becomes
the regular trace condition. The divergence constraint for on-shell massive tensors then follows from
$D\cdot \partial_U \ \Phi=0$. This story is unaltered by inclusion of a fixed homogeneity
\begin{equation}\label{hom}
\big(U\cdot \partial_U - X\cdot \partial_X \big) \, \Phi(X,U)=\t\,  \Phi(X,U)\, ,
\end{equation}
which together with the gauge condition~$\partial_{M}\partial^M\, \Phi(X,U)=0$ give eigenvalues for the Laplacian and thus masses
for fields along~$\cM$ according to the Weyl-weight relationship~\eqref{massweight}~\cite{Gover:2008pt,Gover:2008sw}.
The above gauge fixing gives us a clear link between the~$(d+1)$-dimensional projective construction of~\cite{Biswas:2002nk,Hallowell:2005np} and its tractor formulation, 
which can be viewed as its~$(d+2)$-dimensional lift.

We will also need an integration formula based on the~$(d+2)$-dimensional ambient measure:
\begin{equation*}
	\int_{\sigma}\,\cL\ := \int \frac{d^{d+2} X\sqrt{G}\ \delta(X^{2})}{
	{\rm vol}(GL(1))\,\sigma(X)^{d}}\,\cL\,.
\end{equation*}
This formula deserves quite some explanation: $\cL$ stands for any scalar, ambient function of vanishing homogeneity. The ambient metric determinant~$\sqrt{G}$ has homogeneity~$d+2$ 
(of course, for the conformally flat case in standard coordinates, it is unity).  The delta function of the cone constraint has homogeneity
$-2$ and the factor~$\sigma(X)^{-d}$ has homogeneity~$-d$,  thus the  integral inside the square brackets has zero conformal weight and corresponds to a Weyl invariant integral in~$d$-dimensions. The delta function removes one coordinate (complementary to the cone) and therefore leaves an integral over the projective cone
of a projectively invariant quantity. Thus the result is proportional to the volume of the dilation group, denoted by vol($GL(1)$), multiplied by a Weyl invariant~$d$-dimensional integral describing the physics we are interested in. Such integrals have been utilized in various contexts, see for example~\cite{Gover:2009vc,Bars:2006dy,Bars:2008sz,Bars:2010xi,Bonezzi:2010jr}. The last step 
 is to extract an integral over the actual constant curvature space where our theory lives.
The point is simply (as discussed earlier) that a Weyl invariant quantity~$I[g,\sigma]=I[\Omega^2\,g,\Omega\,\sigma]$ with a St\"uckelberg
shift symmetry ultimately encodes a canonical (pseudo-)Riemmanian one obtained by choosing a gauge~$\sigma(X)=1$\,. 

\subsection{Harmonic gauge vertex}

The result~\eqref{tractor cons} was actually not unexpected. Indeed, one can recover from the 
ambient approach in the harmonic gauge.  In this case following~\cite{Joung:2011ww} the general ansatz for the transverse-traceless part of the cubic interaction in the flat~$(d+1)$-dimensional projective space can be written in terms of harmonic gauge-fixed tractors as
\begin{equation*}
S^{(3)}\,\approx\,\left\langle \s^{\sum_i \tau_i}\,C(Y_i,Z_i)\right\rangle_{\Phi_{1}\Phi_{2}\Phi_{3}}\, ,
\end{equation*}
where
$Z_i\,:=\,\Z_i$ and $Y_i\,:=\,\partial_{U_i}\cdot\partial_{X_{i+1}}$.
Notice that no normal ordering is required since reorderings produce only~$I\cdot\partial_X$ and~$I\cdot\partial_U$.
To compute the gauge variation, we need an integration by parts formula for
$
\left\langle\,\s^{\sum_i \tau_i}\,\partial_{X^M}\left(\,\cdots\right)\right\rangle\, 
$,
where all derivatives acting on the delta-function measure are encoded by
$
\left\langle\,\s^{\sum_i \tau_i}\,\partial_{X^M}\left(\,\cdots\right)\right\rangle \,\approx\,-\,\left\langle\,\s^{\sum_i \tau_i}\,X^M\,\hat\g\,\left(\,\cdots\right)\right\rangle
$
and~$\hat\g$ is defined in~\eqref{dhat}.
Thus we have two main identities; firstly:
\[
\left\langle\,\s^{\sum_i \tau_i}\partial_{X}\cdot\partial_{U_i}\left(\,\cdots\right)\right\rangle\approx-\left\langle\,\s^{\sum_i \tau_i}\,\hat\g\,X_i\cdot\partial_{U_i} \left(\,\cdots\right)\right\rangle
\approx\left\langle\,\s^{\sum_i \tau_i}\,\hat\g\,Z_{i+1}\partial_{Y_{i-1}} \left(\,\cdots\right)\right\rangle\,.
\]
And second:
\ba\nonumber
&&\left\langle\,\s^{\sum_i \tau_i}\,\partial_{X}\cdot\partial_{X_i}\left(\,\cdots\right)\right\rangle \approx -\left\langle\,\s^{\sum_i \tau_i}\,\hat\g\,X_i\cdot\partial_{X_i}\left(\,\cdots\right)\right\rangle\nn
&&\approx\, -\left\langle\,\s^{\sum_i \tau_i}\,\hat\g\,\left[Y_i\partial_{Y_i}-Y_{i-1}\partial_{Y_{i-1}}+ Z_{i-1}\partial_{Z_{i-1}}+Z_{i+1}\partial_{Z_{i+1}}-\t_i+2\right] \left(\,\cdots\right)\right\rangle\,.\nonumber
\ea
We can now impose  gauge consistency  with respect to the massless field~$\Phi_1$:
\begin{equation*}
\delta^{(0)}_{E_1}S^{(3)}\,\approx \,\left\langle\,\s^{\sum_i \tau_i}\left[\partial_{X_1}\!\cdot\partial_{X_2}\,\partial_{Y_1}\,+\, \partial_{U_2}\cdot\partial_X\,\partial_{Z_3}\,+\,Y_3\partial_{Z_2}\,-\,Y_2\partial_{Z_3}\right]\,C(Y_i,Z_i)\right\rangle_{E_1\Phi_{2}\Phi_{3}}\, .
\end{equation*}
Using the above identities together with the on-shell relation
$
\partial_{X_1}\!\cdot\partial_{X_2}\approx\tfrac{1}{2}\,\partial_{X}\cdot(\partial_{X_1}+\partial_{X_2}-\partial_{X_3})
$,
one then gets
\begin{equation*}
\delta^{(0)}_{E_1}S^{(3)}\,\approx \,\left\langle\,\s^{\sum_i \tau_i}\left[Y_3\partial_{Z_2}-Y_2\partial_{Z_3}\, -\,\hat\g\left(Y_3\partial_{Y_3}-Y_2\partial_{Y_2}+\tfrac{\tau_2-\tau_3}2\right)\partial_{Y_1} \right]\,C(Y_i,Z_i)\right\rangle_{E_1\Phi_{2}\Phi_{3}}\, .
\end{equation*}
Thus, we finally recover the differential equation for~$C(Y_i,Z_i)$:
\begin{equation*}
\left[Y_3\partial_{Z_2}-Y_2\partial_{Z_3}\, -\,\hat\g\left(Y_3\partial_{Y_3}-Y_2\partial_{Y_2}+\tfrac{\tau_2-\tau_3}2\right)\partial_{Y_1} \right]\,C(Y_i,Z_i)=0\, .\label{harmonic}
\end{equation*}
In fact, this is exactly the same equation as obtained in Section~\ref{vgi}.

\bibliographystyle{JHEP}
\bibliography{ref}

\end{document}